\documentclass[prd,reprint,superscriptaddress,nofootinbib,longbibliography,aps]{revtex4-1}

\usepackage{graphicx}
\usepackage{longtable}
\usepackage{amsmath}
\usepackage{color}
\usepackage{amssymb}
\usepackage{subfigure}
\usepackage{tabularx}
\usepackage[colorlinks=true]{hyperref}
\usepackage{cleveref}
\usepackage[normalem]{ulem}
\usepackage{slashed}
\usepackage[toc]{appendix}
\usepackage{bm}
\usepackage{braket}
\usepackage{comment}
\usepackage[dvipsnames]{xcolor}   
\usepackage{orcidlink}

\begin{document}

\title{Realization Variance of Gravitational Wave Background Anisotropies from Shot Noise for Pulsar Timing Arrays}

\author{Meng-Xiang Lin\,\orcidlink{0000-0003-2908-4597}}
\altaffiliation{CITA National Fellow}
\email{mengxiang\_lin@sfu.ca}
\affiliation{Department of Physics, Simon Fraser University, Burnaby, British Columbia, V5A 1S6, Canada}
\affiliation{Canadian Institute for Theoretical Astrophysics (CITA), University of Toronto, 60 St George Street, Toronto, Ontario M5S 3H8, Canada}

\author{Adam Lidz\,\orcidlink{0000-0002-3950-9598}}
\affiliation{Center for Particle Cosmology, Department of Physics and Astronomy, University of Pennsylvania, Philadelphia, Pennsylvania 19104, USA}

\author{Chung-Pei Ma\,\orcidlink{0000-0002-4430-102X}}
\affiliation{Department of Astronomy, University of California, Berkeley, California 94720, USA}
\affiliation{Department of Physics, University of California, Berkeley, California 94720, USA}

\begin{abstract}
Shot-noise anisotropies in the nHz gravitational wave background (GWB) are a promising target for pulsar timing arrays (PTAs). If the nHz GWB is sourced by merging supermassive black hole binaries (SMBHBs), as current evidence suggests, the shot-noise signal is expected to be large, potentially of order unity at observing frequencies of $f \sim 1 \, \mathrm{yr}^{-1}$. In this regime, the signal is dominated by rare bright binaries, and Poisson fluctuations in the discrete SMBHB population produce significant spatial anisotropies. 
Here, we use Monte Carlo simulations to model the realization-to-realization scatter in the shot-noise, sampling from empirically calibrated models of the SMBHB source populations. We find that the probability distribution of shot-noise amplitudes is broad, spanning a factor of $\sim 50$ (95\% interval) at fixed frequency, with a long tail towards high amplitudes. The most probable and median amplitudes lie significantly below the ensemble means by factors of $\sim 2-3$, implying that the shot-noise in typical realizations is smaller than the mean. The ensemble-averaged shot-noise also differs from simple estimates based on moments of the strain, $\langle h^4 \rangle/\langle h^2 \rangle^2$,  because the average of a ratio is not equal to the ratio of the averages (i.e., $\langle X/Y \rangle \ne \langle X \rangle/\langle Y \rangle$). This difference is a factor of $\sim 3$ at $f = 0.1 \, \rm{yr}^{-1}$, growing to larger than two orders of magnitude by $f \sim 1 \, \rm{yr}^{-1}$, where the GWB is dominated by low abundance, high-strain sources.
Shot-noise nevertheless provides a powerful diagnostic for understanding the GWB and SMBHB populations; interpreting PTA measurements, however, requires modeling its full probability distribution. 
\end{abstract}

\maketitle

\section{Introduction}
\label{sec:intro}

Pulsar Timing Arrays (PTAs) have recently reported measurements of the gravitational wave background (GWB) at nHz frequencies~\cite{NANOGrav:2023gor,Xu:2023wog,EPTA:2023fyk,Reardon:2023gzh,InternationalPulsarTimingArray:2023mzf,Miles:2024seg}. The GWB at these frequencies is most likely sourced by inspiraling supermassive black hole binaries (SMBHBs), although subdominant or alternative contributions remain viable (see e.g. recent review articles~\cite{Taylor:2025lxy,Sesana:2025udx,Mingarelli:2026pur}). If the PTA signal indeed mainly traces SMBHBs, it offers a new observational anchor for understanding the growth and evolution of SMBH populations, their connections to galaxy formation, and the dynamical processes that drive black hole mergers. 

A key next step is to develop new analysis strategies to extract additional information from current and upcoming data. One highly anticipated approach is to measure spatial anisotropies in the GWB \citep{Mingarelli:2013dsa,Taylor:2013esa,Taylor:2015udp,Roebber:2016jzl,Taylor:2020zpk,Ali-Haimoud:2020ozu,Ali-Haimoud:2020iyz,NANOGrav:2023tcn}. Spatial anisotropies in the GWB imprint distinctive departures from the Hellings-Downs \citep{Hellings:1983fr} angular correlations between the timing residuals towards pulsar pairs. Measuring these may allow PTAs to move beyond determining only the global-average GWB monopole signal. 
Although GWB anisotropy measurements will likely be limited to a small handful of low-$\ell$ (large angular scale) modes for the foreseeable future \citep{NANOGrav:2023tcn}, such measurements should still be highly informative. In particular, the shot-noise signal in merging SMBHB scenarios is expected to be sizable, especially at the high frequency end of the PTA band (e.g., at $f \gtrsim 10-20$ nHz), where only a few sources may dominate the GWB, leading to strong Poisson fluctuations across the sky~\citep{Lin:2026owz}.
Although previous work~\cite{Sato-Polito:2023spo,Lin:2026owz} characterized the mean shot-noise signal with an analytical approach, a complete treatment requires modeling the full realization-to-realization probability distribution function (PDF). In the regime of interest, the shot-noise amplitude can vary substantially from realization to realization. 

Once measurable, the shot-noise signal is expected to provide additional information about the merging SMBHB populations, beyond that encoded in the GWB monopole \citep{Sato-Polito:2023spo,Lin:2026owz}. For example, the GWB shot-noise amplitude depends on the fourth moment of the source strain distribution and therefore receives more weight from the high-mass tail of the SMBH mass function \citep{Lin:2026owz}. Likewise, bright nearby sources can provide potentially dominant shot-noise contributions. Finally, the shot-noise signal depends on the source ``residence times'', which statistically characterize the amount of time members of the merging SMBHB population spend radiating gravitational waves in a given frequency interval. 

Recent work on the GWB monopole has emphasized the importance of modeling the full realization-to-realization PDF \citep{Sesana:2008mz,Becsy:2022pnr,Agazie:2024jbf,Sato-Polito:2024lew,Lamb:2024gbh,Konstandin:2024fyo,Ali-Haimoud:2026sbk}. At least in conventional SMBHB merger scenarios, the characteristic strain in individual realizations scatters widely, with a long tail towards high strain amplitudes. The width and skewness of this distribution grow strongly with frequency as one moves from a signal with many contributing sources to the low source count regime. Therefore, PTA parameter inference efforts must account for the full characteristic strain PDF across sky realizations in each model of interest, rather than relying solely on ensemble-averaged predictions. 
Furthermore, the frequency-to-frequency scatter in PTA measurements probes independent Poisson draws from the characteristic strain PDF, providing a further diagnostic of the source abundance \citep{Sato-Polito:2024lew}.

The realization-to-realization scatter considered here should be distinguished from the ``cosmic variance'' of the Hellings-Downs correlation discussed in the PTA literature \cite{Allen:2022dzg,Allen:2024mtn,Konstandin:2024fyo}, which refers to the scatter of the measured pulsar-pair correlations around the ensemble-mean Hellings-Downs curve, sourced mainly by interference between GW sources
radiating at overlapping frequencies for a fixed source population. 
That effect persists even in the Gaussian, many-source limit, in close analogy with Cosmic Microwave Background cosmic variance. 
In contrast, the realization variance studied here is Poisson scatter in the discrete SMBHB population itself --- the analogue of shot noise in galaxy surveys --- and vanishes in the many-source limit. 
Both effects are irreducible for our single sky at a fixed frequency, and both should enter the interpretation of PTA anisotropy measurements; notably, the source-discreteness correction to the Hellings-Downs cosmic variance \cite{Allen:2024mtn} is controlled by the same fourth-moment-weighted effective source count, $N_{\rm eff}$, that sets the shot-noise level studied in this work.

In this work, we investigate the previously uncharacterized realization-to-realization scatter in the shot-noise anisotropy level. As with the characteristic strain field, the shot-noise has a broad, non-Gaussian, and positively skewed PDF. We use Monte Carlo simulations of merging SMBHB populations, drawn from simple yet flexible and empirically calibrated models \citep{Liepold:2024woa,Sato-Polito:2023gym}, to characterize the full PDFs of the shot-noise as well as the second and fourth moments (see also \cite{Lamb:2024gbh}) of the strain field. 
We investigate their dependence on SMBHB merger models, the sensitivity to nearby sources, and the impact of an alternative residence time model. Finally, we quantify how the shot-noise and fourth moment distributions narrow after conditioning on measurements of the characteristic strain. 

Specifically, Sec.~\ref{sec:stats} introduces the relevant statistical properties of the GWB, anticipates some aspects of the Monte Carlo simulation results, and describes the adopted SMBHB merger models. Sec.~\ref{sec:mc_results} presents results from our Monte Carlo simulations, while Sec.~\ref{sec:model_dependence} discusses the dependence on a few key ingredients of the merger models. Finally, we summarize our conclusions in Sec.~\ref{sec:conclusions}.

\section{Strain-Field Statistics and SMBHB Merger Models}
\label{sec:stats}
First, we review the statistical properties of the GWB strain fields. We generally follow the notation and conventions of \cite{Lin:2026owz}. We then provide an overview of our Monte Carlo simulation methodology, and briefly summarize the fiducial empirically calibrated SMBHB merger model adopted in this work \cite{Liepold:2024woa}.

\subsection{Shot-noise and Moments of the Characteristic Strain Field}
\label{sec:strain}
We aim to characterize the probability distribution of shot-noise anisotropy levels across different realizations of the sky.
Following \cite{Lin:2026owz}, we generalize the characteristic strain field \cite{Phinney:2001di} to allow spatial fluctuations in this quantity:
\begin{equation}
\delta_{h^2}(f, \boldsymbol{\Omega}) = \frac{h^2(f, \boldsymbol{\Omega}) - \langle h^2 \rangle_\Omega}{\langle h^2 \rangle_\Omega},
\end{equation}
where $h^2(f, \boldsymbol{\Omega})$ is the characteristic strain-squared per steradian in direction $\boldsymbol{\Omega}$ for a logarithmic interval $d \ln f$ around observed frequency $f$, and $\langle  \rangle_\Omega$ denotes the angle-average across the sky. An important distinction for this work is that angle-averaged quantities in a single sky realization generally differ from ensemble averages across many realizations.
We denote estimates from a particular realization with a ``hat'' notation henceforth so that $\hat{h}^2_c(f) = 4 \pi \langle \hat{h}^2 \rangle_\Omega$ is the characteristic strain in an individual realization,
likewise the fourth-moment $\hat{h}^4_c(f) = 4\pi \langle \hat{h}^4 \rangle_\Omega$.

The shot-noise anisotropy for the $\delta_{h^2}(f, \boldsymbol{\Omega})$ field
in a particular realization is 
\begin{equation}
    \hat{C}_{\rm shot} = \frac{\langle \hat{h}^4 \rangle_\Omega}{\left[\langle \hat{h}^2 \rangle_\Omega \right]^2},
\end{equation}
which carries steradian units, and is independent of $\ell$. Given these conventions, our $\hat{C}_{\rm shot}/(4 \pi)$ is equivalent to the NANOGrav
shot-noise quantity \cite{NANOGrav:2023tcn}, $\tilde{C}_{\ell > 0}/\tilde{C}_{\ell=0}$, where the tildes indicate NANOGrav's definition based on the power spectrum of $h^2(f,\boldsymbol{\Omega})$, rather than $\delta_{h^2}(f,\boldsymbol{\Omega})$  (see \cite{Lin:2026owz} for further details). 

Next, we would like to connect the shot-noise and related moments of the characteristic strain field to models for SMBHB mergers following references
\citep{Phinney:2001di, Sesana:2008mz, Sato-Polito:2023spo, Sato-Polito:2023gym, Liepold:2024woa,Lamb:2024gbh, Lin:2026owz}.
For this purpose, it is useful to express the ensemble-averaged characteristic strain per logarithmic frequency interval as:
\begin{equation}
h^2_c(f) = \int dM_{\mathrm{BH}} dq dz \, \frac{d^4N}{dM_{\mathrm{BH}} dq dz d\ln f} \, h^2_s(M_{\mathrm{BH}},q,z,f)\,,
\label{eq:h2}    
\end{equation}
where $h^2_s(M_{\mathrm{BH}},q,z,f)$ is the strain amplitude per source given the physical quantities (total black hole mass $M_{\mathrm{BH}}\equiv M_1+M_2$, mass ratio $q\equiv M_1/M_2\in [0,1]$, redshift $z$, and frequency $f$), and $\frac{d^4N}{dM_{\mathrm{BH}} dq dz d\ln f}$ is the ensemble-averaged merger distribution. The characteristic strain can be simply calculated by summing up the products of the number of merging sources and the strain per source.
This expression is equivalent to an alternative formula which adds up the cumulative energy emitted in GWs \cite{Phinney:2001di}. 

We further assume that the number of mergers per logarithmic frequency interval can be factorized ~\cite{Sato-Polito:2023gym,Liepold:2024woa}:
\begin{equation}
\frac{d^4N}{dM_{\mathrm{BH}} dq dz d\ln f} = p_z(z) p_q(q) \frac{dn}{dM_{\mathrm{BH}}} \frac{dV_\mathrm{c}}{dz} \frac{dz}{dt_r}\frac{dt_r}{d\ln f}\,.
\label{eq:d4N}
\end{equation}
Here, $p_z(z)$ describes the redshift distribution of SMBHB mergers, and $p_q(q)$ gives their mass-ratio distribution. Both distributions are normalized to integrate to unity. 
The SMBH mass function, $dn/dM_{\mathrm{BH}}$, describes the abundance of SMBHs per comoving volume per unit black hole mass. 
The comoving volume and source-frame time are denoted by $V_\mathrm{c}$ and $t_r$, respectively. The factor $dt_r/d\ln f$ describes how much source-frame time an SMBHB spends emitting gravitational waves in a given logarithmic frequency interval, referred to as the residence time. Assuming purely GW-driven inspiral and circular orbits, this is given by:
\begin{equation}\label{eq:dfdt}
    \frac{d\ln f_r}{dt_r} = \frac{96}{5}\pi^{8/3} \left(\frac{G\mathcal{M}}{c^3} \right)^{5/3}f_r^{8/3},
\end{equation}
where $f_r = f(1+z)$ is the source-frame frequency, and $\mathcal{M}\equiv M_{\rm BH}[q/(1+q)^2]^{3/5}$ is the chirp mass.

Finally, the inclination-angle and polarization-averaged strain per source, assuming circular orbits, is \citep{Sesana:2008mz,Sato-Polito:2023gym}
\begin{eqnarray}
    h^2_s(M_{\mathrm{BH}},q,z,f) &=& \frac{32 \pi^{4/3}}{5 \,c^{8}}  \left(G M_{\mathrm{BH}}\right)^{10/3} \frac{q^2}{(1+q)^4} \nonumber \\ 
        &\times& \frac{(1+z)^{10/3}}{d^2_L(z)} f^{4/3},
\label{eq:h2_source}
\end{eqnarray}
where $d_L(z)$ is the luminosity distance. For simplicity, we adopt this inclination-angle and polarization-averaged expression throughout. In reality, individual binaries should sample a distribution of orientations, introducing an additional source of variance beyond that considered here. Quantifying the impact of inclination angle variations on the strain and shot-noise distributions is left for future work.

Similarly to Eq.~\eqref{eq:h2}, the ensemble-averaged fourth moment of the strain field can be calculated as:
\begin{equation}
h^4_c(f) = \int dM_{\mathrm{BH}} dq dz \, \frac{d^4N}{dM_{\mathrm{BH}} dq dz d\ln f} \, h^4_s(M_{\mathrm{BH}},q,z,f)\, .
\label{eq:h4}
\end{equation}
Eqs.~\eqref{eq:h2} and \eqref{eq:h4} are equivalent to the results given in Sec.~II.B of~\citep{Lin:2026owz}, but are written explicitly in terms of strain amplitude per source $h^2_s(M_{\rm BH}, q, z, f)$, which is convenient for our Monte Carlo simulations. 
Unless stated otherwise, both $h^2_c(f)$ and $h^4_c(f)$ are ensemble-averaged quantities computed from a merger model, while $\hat{h}^2_c(f)$ and $\hat{h}^4_c(f)$ represent particular realizations. 

\subsection{Generating Monte Carlo Realizations}
\label{sec:mc_method}
We then generate realizations of the strain moments of Eqs.~\eqref{eq:h2} and \eqref{eq:h4} by sampling from a Poisson distribution:
\begin{eqnarray}
\label{eq:h2_poisson_sampling}
\hat{h}^2_c(f) &=& \sum_i N_i(f) \, h^2_i(f)\,,\\
\label{eq:h4_poisson_sampling}
\hat{h}^4_c(f) &=& \sum_i N_i(f) h^4_i(f)\,,
\end{eqnarray}
where $i$ is an index that labels cells in $M_{\mathrm{BH}}$, $q$, and $z$, and $N_i$ is a random integer drawn from the Poisson probability distribution of SMBHB mergers occurring in cell $i$, $\mathcal{P}(N_i | \overline{\Delta N_i})$, given an ensemble average expected merger rate 
\begin{equation}
    \overline{\Delta N_i}(f) = \frac{d^4N}{dM_{\mathrm{BH}} dq dz d\ln f} \Delta M_{\mathrm{BH}} \Delta q \Delta z\,.
    \label{eq:dn_avg}
\end{equation}
$h^2_i(f)$ is the individual source strain amplitude in cell $i$. The source strain depends on the cell properties following Eq.~\eqref{eq:h2_source}.
We emphasize that $\overline{\Delta N_i}(f)$ and all derived quantities are defined per $d \ln f$ rather than within discrete frequency bins. 
Thus, $\hat{h}^2_c(f)$ denotes a realization of the characteristic strain field at frequency $f$.

The shot-noise anisotropy level from a particular Monte Carlo realization is then determined from the ratio of these quantities, as:
\begin{equation}
\frac{\hat{C}_{\mathrm{shot}}}{4 \pi}(f) = \frac{\hat{h}^4_c(f)}{[\hat{h}^2_c(f)]^2}= \frac{\sum_i N_i h^4_i}{\left[ \sum_i N_i h^2_i\right]^2}\,,
\label{eq:cshot_poisson_sampling}
\end{equation}
where we omit the frequency dependence in the right-hand side of the second equality and in the discussion below for brevity.
Note that in our Monte Carlo simulations, we evaluate this quantity directly from realizations of the sky-averaged moments, rather than constructing full angular maps. This is sufficient because the shot-noise is $\ell$-independent and its amplitude is determined by these moments. 
When comparing to PTA measurements, the differential quantities calculated in our Monte Carlo simulations must be averaged over finite frequency bins incorporating the weighting schemes used in the data analysis (see \S~\ref{sec:conclusions}).

From these expressions, we can already anticipate a key feature of the simulation results presented in the following section. Specifically, the ensemble-averaged shot-noise, $\langle \hat{C}_\mathrm{shot}/(4 \pi)\rangle$, is not the same as the ratio $\langle \hat{h}^4_c \rangle/\langle \hat{h}^2_c\rangle^2$.\footnote{Here the averaging indicates an ensemble average, distinct from the angle averages across a particular sky realization, which are denoted as e.g. $\langle h^2 \rangle_\Omega$.} This is an example of the fact that $\langle X/Y \rangle \neq \langle X \rangle/\langle Y\rangle$ for non-linear functions. Here, the shot-noise involves non-linear functions of the strain fields, leading to departures between its ensemble average and the ratio of the moments involved. 

A key distinction between these two quantities is that, whereas $\langle\hat h^4_c\rangle/\langle\hat h^2_c\rangle^2$ can be arbitrarily large~\citep{Lin:2026owz}, $\hat{C}_{\rm shot}/4\pi$ (and hence its ensemble average) is strictly bounded above by unity. To prove this, define the weight of each cell as
\begin{equation}
    w_i \equiv \frac{h^2_i}{\sum_j N_j h^2_j}\,,
\end{equation}
so that we can rewrite Eq.~\eqref{eq:cshot_poisson_sampling} as
\begin{equation}
    \frac{\hat{C}_{\mathrm{shot}}}{4 \pi} = \sum_i N_i w_i^2\,.
\end{equation}
By definition, we have $\sum_i N_i w_i = 1$. 
Since $0<w_i\le 1$ for occupied cells, we obtain the upper bound
\begin{equation}
    \frac{\hat{C}_{\mathrm{shot}}}{4 \pi} \le \sum_i N_iw_i =1\,,
\end{equation}
where equality holds only when there is a single source $N\equiv\sum_i N_i=1$.

On the other hand, from the mean inequality $\sqrt{\frac{\sum_k x_k^2}{N}}\ge\frac{\sum_k x_k}{N}$ and taking $x_k$ as the weights of individual sources, we have 
\begin{equation}\label{eq:Cshot_lower}
    \frac{\hat{C}_{\mathrm{shot}}}{4 \pi} \ge N\left(\frac{\sum_i N_i w_i}{N}\right)^2 =\frac{1}{N}\,,
\end{equation}
where equality holds only when all sources have the same strain amplitude. 
Therefore, it is natural to interpret shot-noise as the inverse of the effective number of sources $\hat{C}_\mathrm{shot}/(4 \pi)\equiv1/N_{\mathrm{eff}}$~\cite{Lin:2026owz}, and it is larger when a few bright sources dominate the GWB.
The only caveat for Eq.~\eqref{eq:Cshot_lower} is when a realization contains no sources, and in this case $\hat{C}_{\rm shot}=0$.
In practice, the SMBH mass function rises steeply towards low masses, and the residence times are longer at low mass. Consequently, within the frequency range considered here ($f \lesssim 30$ nHz), our Monte Carlo realizations always contain at least some sources even though the low-mass strain contributions may be very small.

Both $\hat{h}^4_c$ and $\hat{C}_{\rm shot}/4\pi$ could be direct observables depending on the experimental design and analysis method. We present the simulation results of both quantities in the next section. The current NANOGrav anisotropy analyses~\citep{NANOGrav:2023tcn} directly probe $\hat{C}_{\rm shot}$, but in principle $\hat{h}^4_c$ could also be obtained using a different analysis method.

In the single source limit, the pulsar-pair correlations in an individual realization follow a well-defined overlap reduction function, distinct from the Hellings-Down curve~\citep{Mingarelli:2026kjw}. This can be used to complement traditional anisotropy searches at high frequencies.

\begin{figure*}
    \centering
    \includegraphics[width=0.99\linewidth]{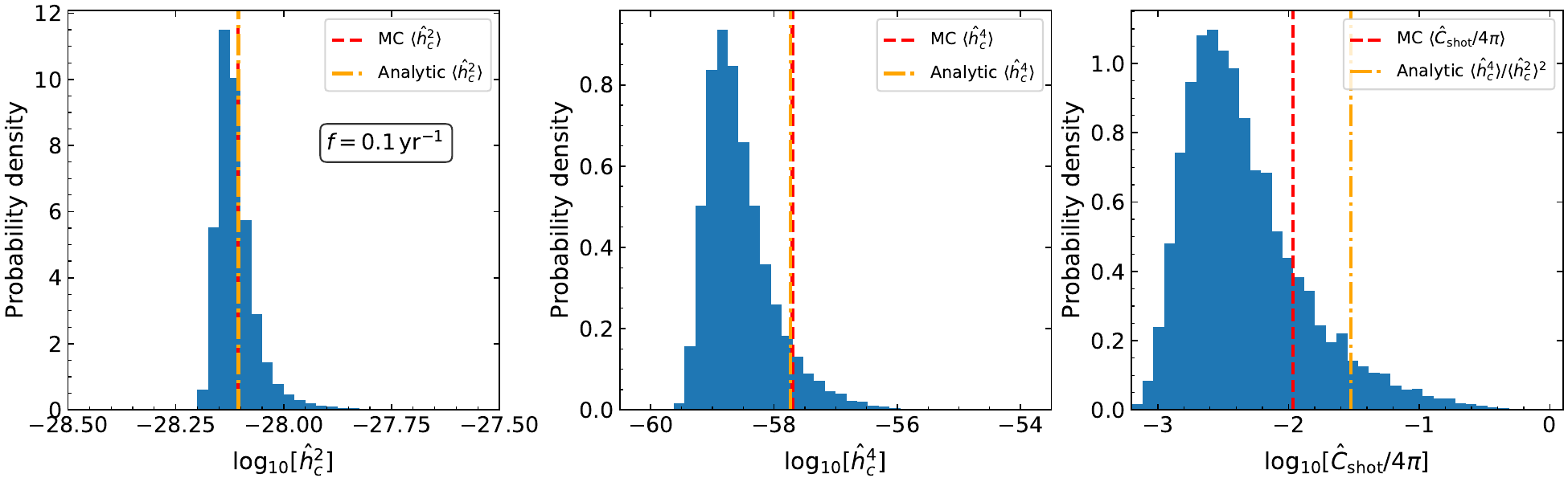}
    \includegraphics[width=0.99\linewidth]{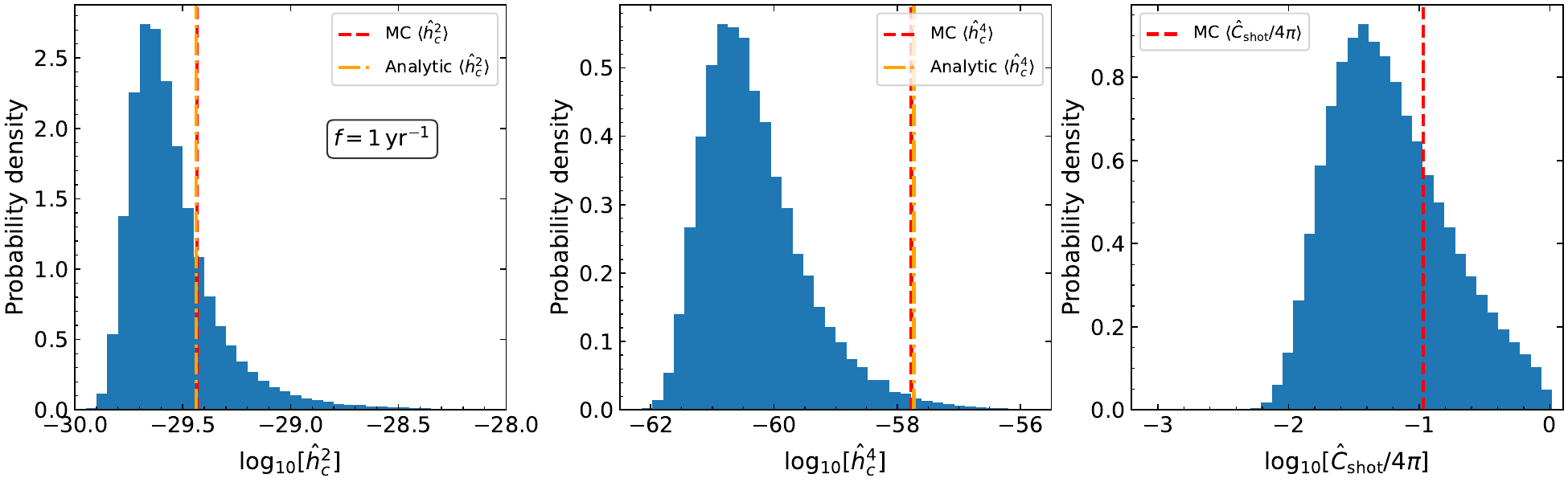}
    \caption{Probability distribution of the strain amplitude $\hat{h}^2_c$ (left), fourth moment $\hat{h}^4_c$ (middle), and shot-noise amplitude $\hat{C}_{\mathrm{shot}}/(4 \pi)$ (right) from Monte Carlo simulations at two frequencies, $f =0.1 \,\mathrm{yr}^{-1}$ (top) and $f = 1 \,\mathrm{yr}^{-1}$ (bottom). 
    In each panel, the red-dashed line indicates the mean value estimated from the Monte Carlo realizations, while the orange dot-dashed line is an analytic prediction for the ensemble average value. Each distribution has a long tail towards large field values,  reflecting contributions from rare bright sources, and the peak and median values are smaller than the ensemble averages. The Monte Carlo simulations reproduce the expected mean value for $\hat{h}^2_c$ and $\hat{h}^4_c$ but the mean shot noise from the simulations differs from the naive analytic expectation of $\langle \hat{h}^4_c \rangle/\langle \hat{h}^2_c \rangle^2$, as discussed in the text. (The analytic result is not shown in the lower right panel as it lies to the right of the shot-noise range included along the x-axis.) 
    } 
    \label{fig:strain_statistics_threepanel}
\end{figure*}

\subsection{SMBHB Merger Models}
\label{sec:merger_models}
Our fiducial model for the SMBHB merger rate follows that of \cite{Liepold:2024woa}, hereafter LM24. In this model, the SMBH mass function is calculated as a convolution between the galaxy stellar mass function and a lognormal probability distribution describing the correlation between black hole mass and stellar mass. The redshift and mass-ratio distributions of mergers follow
\begin{equation}\label{eq:pz_pq}
    p_z(z) \propto z^\gamma e^{-(z/z_*)^2}\,,\quad p_q(q) \propto q^2\,
\end{equation} 
with $\gamma=1.0$, and $z_*=0.5$. The proportionality constants are determined so that each distribution integrates to unity.\footnote{In the case of $p_z(z)$ the normalization is determined after imposing the sharp $z_{\rm min}$ cut-off discussed below.} 

As discussed in \cite{Lin:2026owz}, the ensemble-averaged value
of $h^4_c(f)$ in Eq.~\eqref{eq:h4} is formally divergent in this model as $z \rightarrow 0$. This divergence reflects the dominant influence of rare nearby sources to the higher moments of the strain fields.
In addition, $h^4_c(f)$ is sensitive to the high-mass tail of the mass function. Our fiducial calculations therefore truncate $p_z(z)$ sharply below $z < z_{\rm min} = 0.05$, and the SMBH mass function above
$M_{\rm BH, max} = 10^{10.5} M_\odot$, comparable to the most massive BHs observed thus far \citep{Liepold25}.

An alternate SMBHB merger model is that of \cite{Sato-Polito:2023gym}, hereafter SZQ. This model is based on the observed abundance of galaxies as a function of their velocity dispersion $\sigma$ and correlations between $M_{\rm BH}$ and $\sigma$. The redshift and mass ratio distributions in SZQ also differ:
\begin{equation}\label{eq:pz_pq_szq}
    p_z(z) \propto z^\gamma e^{-(z/z_*)}\,,\quad p_q(q) \propto q^{-1}\,
\end{equation} 
with $\gamma=0.5$, and $z_*=0.3$, where $p_q(q)$ is cut-off below $q_{\rm min} = 0.1$. Compared to LM24, SZQ gives a lower abundance of SMBHs below
$M_{\rm BH} \lesssim 2 \times 10^{10} M_\odot$, but a greater abundance of larger mass SMBHs. However, we truncate the distributions at only slightly larger masses, $M_{\rm BH, max}=10^{10.5} M_\odot$, throughout (see \S~\ref{sec:model_dependence}). The $h^4_c(f)$ values are also more sensitive to $z_{\rm min}$ in SZQ due to the lower $\gamma$ value \cite{Lin:2026owz}. Sec.~\ref{sec:smbh_model} compares the strain-field and shot-noise distributions in LM24 and SZQ and quantifies the $z_{\rm min}$ dependence in LM24.

\section{Simulation Results}
\label{sec:mc_results}
Now that we have established notation, introduced our modeling assumptions, and developed some intuition for what to expect, we generate Monte Carlo simulations of the strain moments according to Eqs.~\eqref{eq:h2_poisson_sampling}-\eqref{eq:cshot_poisson_sampling} and investigate their statistical properties.

\subsection{Strain Moments and Shot-Noise: Distributions and Frequency Dependence}
\label{sec:sn_vs_nu}
Fig.~\ref{fig:strain_statistics_threepanel} shows the resulting probability distributions of the characteristic strain $\hat{h}^2_c$, fourth moment $\hat{h}^4_c$, and shot-noise signals $\hat{C}_{\rm shot}/(4 \pi)$, estimated based on $7 \times 10^5$ Monte Carlo realizations drawn from our fiducial LM24 merger model at each of two representative frequencies, $f = 0.1 \, \mathrm{yr}^{-1}$ (top row) and $f = 1 \, \mathrm{yr}^{-1}$ (bottom row). 
Each distribution is positively-skewed with a long tail towards large values of the strain moments or shot-noise. This effect is most pronounced for $\hat{h}^4_c$ due to its stronger weighting of high-strain sources. 
 
\begin{figure}
\centering
\includegraphics[width=0.99\linewidth]{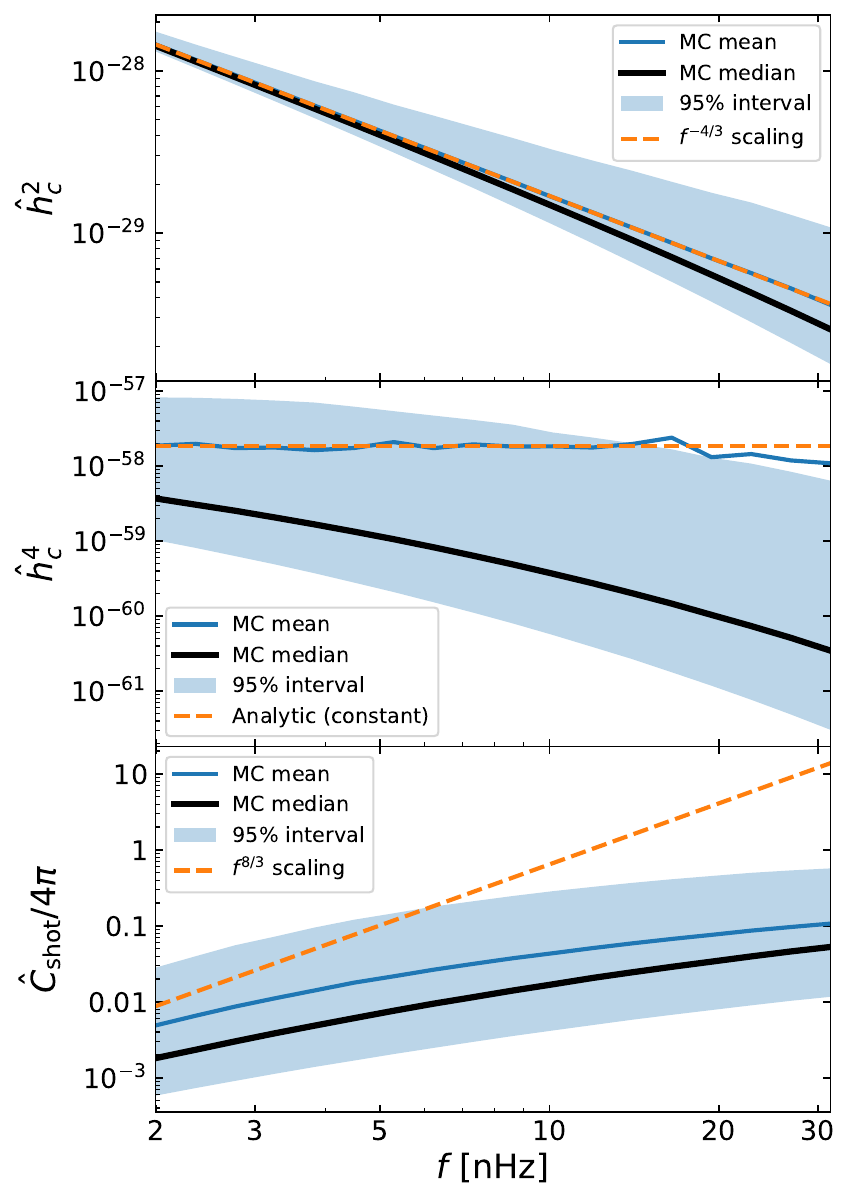}
\caption{Frequency dependence of the characteristic strain $\hat{h}^2_c$ (top), fourth moment $\hat{h}^4_c$ (middle), and shot-noise amplitude $\hat{C}_{\mathrm{shot}}/(4 \pi)$ (bottom). For each quantity, the mean value (blue curve), median value (black curve), and 95\% range of the distribution (blue band) from our Monte Carlo simulations are indicated, and the orange dashed curve represents the analytic prediction for the mean value.
For $\hat{h}^2_c$ and $\hat{h}^4_c$, 
the simulated and analytic mean values agree well and follow the respective analytic frequency scaling. There is, however, a growing difference between the mean and median with frequency, in particular for $\hat{h}^4_c$.
For the shot-noise amplitude in the bottom panel, the moment-based analytic prediction of $\langle \hat{h}^4_c \rangle/\langle \hat{h}^2_c \rangle^2$, which scales as $f^{8/3}$, no longer matches the mean from the Monte Carlo realizations. The discrepancy grows towards high frequencies, as the effective source abundance drops. 
}
\label{fig:frequency_dep}
\end{figure}

Fig.~\ref{fig:frequency_dep} shows the frequency dependence over the full PTA band of $\hat{h}^2_c$ (top), $\hat{h}^4_c$ (middle), and $\hat{C}_{\rm shot}/(4 \pi)$ (bottom) from the same Monte Carlo simulations. In each case, we plot the mean (blue curves) and median (black curves) values, along with the range encompassing 95\% of the realizations (blue bands).  Overall, the distributions broaden towards high frequency because the residence time
scales as $dt_r/d \ln f \propto f^{-8/3}$ (see Eq.~\eqref{eq:dfdt}) in this GW-driven merger scenario. Thus, fewer sources contribute to the high-frequency measurements, and this leads to more Poisson scatter and wider variance in the strain distributions. 
Relatedly, Figs.~\ref{fig:strain_statistics_threepanel} and \ref{fig:frequency_dep} show that the median and most probable values are smaller than the ensemble-averaged results for each distribution. For example, at $f = 0.1 \, \mathrm{yr}^{-1} \approx 3$ nHz, 
the mean and median values of $\hat{h}^2_c$ differ by a factor of $1.03$, while
the median value of $\hat{h}^4_c$ is $8$ times smaller than the mean, and the median shot-noise is $2.9$ times smaller than the mean. 
These differences increase with frequency. At $f = 1 \, \mathrm{yr}^{-1}$, for example, the mean-to-median ratio is $1.4$ for $\hat{h}^2_c$, $440$ for $\hat{h}^4_c$, and $2.0$ for the shot-noise.
The ratio for the shot-noise is slightly smaller than at $f = 0.1 \, \mathrm{yr}^{-1}$ because the peak in the distribution shifts to larger shot-noise amplitudes with increasing frequency, while it is bounded from above by $\hat{C}_{\mathrm{shot}}/(4 \pi) \leq 1$. The distribution is then less positively-skewed, and the distinction between the mean and the median is reduced. 

For both $\hat{h}^2_c$ and $\hat{h}^4_c$ in Figs.~\ref{fig:strain_statistics_threepanel} and \ref{fig:frequency_dep}, the mean values from the Monte Carlo realizations agree with our analytic expectations \cite{Lin:2026owz}, Eqs.~\eqref{eq:h2}-\eqref{eq:h4}.  
This agreement also indicates that the number of Monte Carlo realizations is sufficient for our purposes.
The analytic scaling of $\langle \hat{h}^2_c\rangle \propto f^{-4/3}$ and the frequency independence of $\langle \hat{h}^4_c\rangle$ in the GW-driven regime of SMBHB inspirals are well reproduced in the Monte Carlo simulations.\footnote{The lack of frequency dependence in $\langle \hat{h}^4_c\rangle$ occurs because of a coincidental cancellation between $dt_r/d \ln f \propto f^{-8/3}$ and $h^4_s \propto f^{8/3}$ in the GW-driven scenario; see Eqs.~\eqref{eq:dfdt}-\eqref{eq:h4} or Eq.~(22) in~\cite{Lin:2026owz}.}  
However, the middle panel of Fig.~\ref{fig:frequency_dep} shows that the mean value of the fourth moment is strongly impacted by high-strain outliers, especially at the high frequency end of the PTA band. In fact, by $f \gtrsim 15$ nHz, the average $\hat{h}^4_c$ values fall outside of the 95\% interval of the distribution. Meanwhile, the median $\hat{h}^4_c$ value decreases with increasing frequency as a growing number of realizations lack high-strain sources. 
We therefore caution that the ensemble-averaged value of $\hat{h}^4_c$ gives a significant overestimate of the full $\hat{h}^4_c$ distribution.  

While the analytic predictions for the mean values of $\hat{h}^2_c$ and $\hat{h}^4_c$ agree with those from the Monte Carlo simulations, the prediction for the mean value of $\langle \hat{h}^4_c \rangle/\langle \hat{h}^2_c \rangle^2$ differs from the ensemble mean from the Monte Carlo realizations, $\langle \hat{C}_{\rm shot}/(4 \pi)\rangle$. This follows the trends discussed previously and is illustrated by the difference between the ensemble mean and the moment-based estimate in the right panels of Fig.~\ref{fig:strain_statistics_threepanel} and bottom panel of Fig.~\ref{fig:frequency_dep}.
At $f = 0.1 \, \mathrm{yr}^{-1}$, for example, the moment-based estimate is $2.8$ times larger than the ensemble-mean shot-noise, and this difference is more pronounced at $f = 1 \, \mathrm{yr}^{-1}$, where the moment-based ratio is $130$ times the mean of the Monte Carlo realizations. 
This growing discrepancy between the ensemble-averaged shot-noise and the moment-based estimate with increasing frequency occurs because the moment-based estimate grows without bound as the sources become rare at high frequency, while the shot-noise in each realization, $\hat{C}_{\mathrm{shot}}/(4 \pi)$, must be less than unity. As a result, the Monte Carlo median and mean exhibit a much shallower scaling, approximately $C_{\mathrm{shot}} \propto f^{1.1-1.2}$ (solid black and blue curves in the bottom panel of Fig.~\ref{fig:frequency_dep}), rather than $\propto f^{8/3} \sim f^{2.7}$ (orange dashed line in the same panel) expected from the moment-based calculation in a GW-driven inspiral model. 
These discrepancies reflect a general breakdown of the moment-based estimates in the low source number regime, and underscore the importance of accounting for the full probability distribution of the shot-noise.

Although the mean and median shot-noise amplitudes are smaller than the moment-based estimates, they remain relatively large. For example, the median shot-noise reaches
$\hat{C}_{\rm shot}/(4 \pi) \sim 0.05$ near $f \sim 30 \, \rm{nHz}$ $(1 \, \rm{yr}^{-1})$. This is only a factor of four smaller than the current NANOGrav 95\% upper-bound of $0.20$ \cite{NANOGrav:2023tcn}, although this limit is still primarily from lower frequencies where the PTA survey sensitivity is greatest.  The current NANOGrav bound is also sensitive to the analysis priors adopted.  

These median shot-noise predictions remain much larger than the large-scale structure (LSS) signals predicted in \cite{Lin:2026owz}.
At the low-frequency end near  $f=0.1 \, \rm{yr}^{-1}$, the median shot-noise is $\sim 100$ times larger than the peak LSS signal (near $\ell \sim 10$) for LM24 assuming an optimistic clustering bias of $\langle b_{\rm eff} \rangle = 5 $ \cite{Lin:2026owz}. The ratio of median shot-noise to clustering is similar for the SZQ model (discussed here in \S~\ref{sec:model_dependence}) for $\langle b_{\rm eff} \rangle = 5$. Note that the LSS signal is less dominated by the strain distribution tails, and so the ensemble-averaged LSS signal should be representative, in contrast to the shot-noise.  

Importantly, the width of the shot-noise distribution is broad across the entire frequency band. The 95\% range in shot-noise values at a particular frequency spans a factor of about $\sim 50$ (1.7 dex) in amplitude. 
Interestingly, the fractional width of this band is a relatively weak function of frequency. This may reflect a tradeoff between the increase in shot-noise with frequency, 
and the impact of the $\hat{C}_{\mathrm{shot}}/(4 \pi) \leq 1$ bound.

\begin{figure*}
    \centering
    \includegraphics[width=0.49\linewidth]{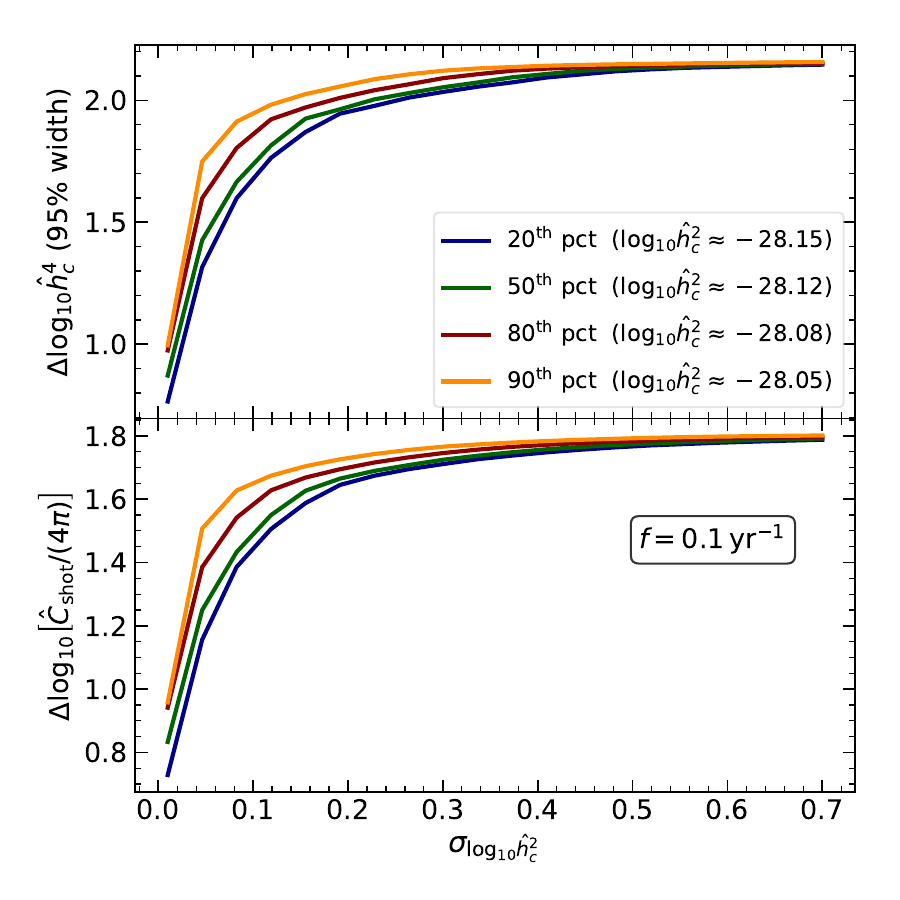}
    \includegraphics[width=0.49\linewidth]{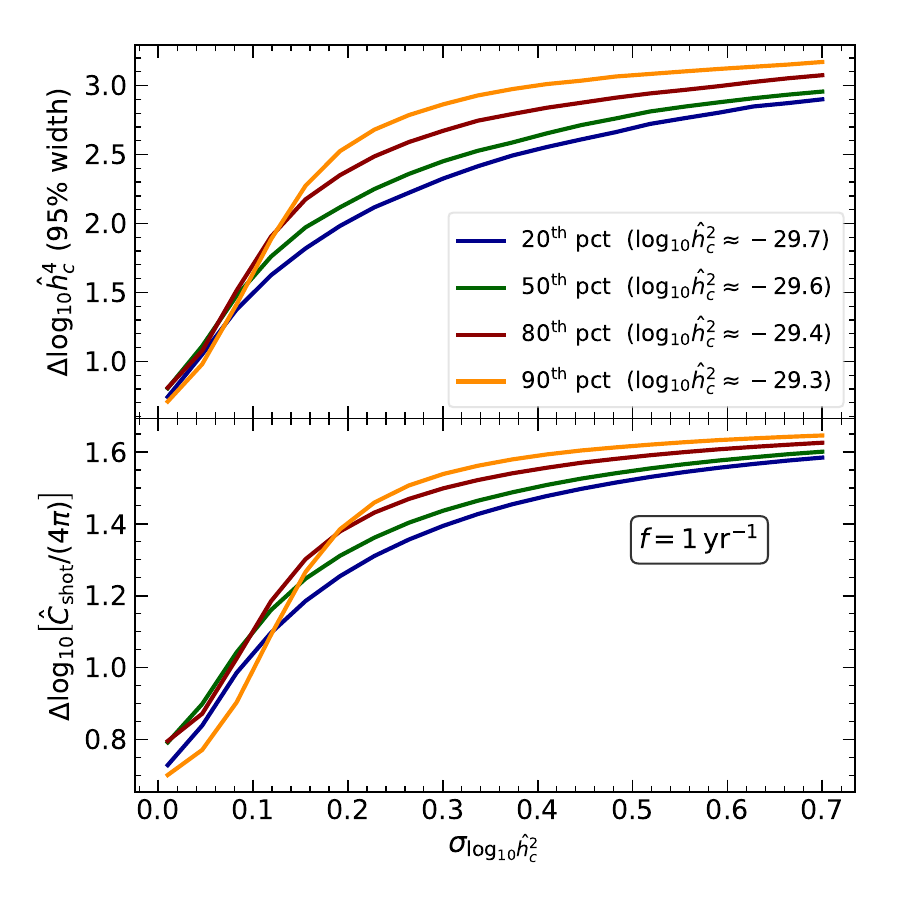}
    \caption{Widths of the conditional distributions of $\hat{h}^4_c$ (top row) and $\hat{C}_{\rm shot}/(4 \pi)$ (bottom row) as a function of the $\hat{h}^2_c$ measurement uncertainties for $f = 0.1 \, \mathrm{yr}^{-1}$ (left) and $f=1 \, \mathrm{yr}^{-1}$ (right). Within each panel, the curves correspond to different assumed central values of $\hat{h}^2_c$, and are labeled by their percentiles within the model distribution. 
    Conditioning narrows the distributions when the measurement error is smaller than the intrinsic scatter in $\hat{h}^2_c$, and becomes ineffective once the measurement error exceeds the intrinsic width.  
    }
\label{fig:conditioned_width}
\end{figure*}    

To assess the dependence of these conclusions on the source abundance, we consider a toy model in which the expected merger rate of Eq.~\eqref{eq:dn_avg} is boosted uniformly by a factor of $50$. In this case, the moment-based mean provides a better description of the Monte Carlo shot-noise results, as expected in the limit of many contributing sources. Still, significant differences remain: the moment-based prediction exceeds the Monte Carlo mean by 20\% at $f \sim 5$ nHz, growing to a factor of $\sim 2$  by $\sim 10$ nHz, and reaching nearly an order of magnitude by $\sim 30$ nHz. Thus, even in this extreme scenario, the moment-based estimate fails at sufficiently high frequencies. We therefore conclude that the ensemble-averaged shot-noise will lie well below the moment-based estimates, especially at high frequencies. 

\subsection{Distributions Conditioned on Background Measurements}
\label{sec:conditioned}

Although $\hat{h}^4_c$ and $\hat{C}_{\rm shot}/(4 \pi)$ show a wide scatter from realization to realization, we note that each of these quantities is correlated with $\hat{h}^2_c$, which is easier to measure. 
Therefore, measurements of $\hat{h}^2_c$ may help reduce the effective width of the $\hat{h}^4_c$ and $\hat{C}_{\rm shot}/(4 \pi)$ distributions, especially with increasing precision in the future.

To explore this, we calculate the PDFs of $\hat{h}^4_c$ and $\hat{C}_{\rm shot}/(4 \pi)$, conditioned on $\hat{h}^2_c$ measurements. 
For notational simplicity, we define $x \equiv \log_{10} \hat{h}^2_c$ and denote either $\log_{10} \hat{h}^4_c$ or $\log_{10} \left[\hat{C}_{\rm shot}/(4 \pi)\right]$ by $y$, working in log-space throughout. We are interested in the distribution of $y$ given a noisy measurement $\tilde{x}$ of $x$.  This conditional distribution is
\begin{equation}
P\left(y | \tilde{x} \right) = \int dx \, P\left(y|x\right) P_{\rm noise}\left(x|\tilde{x}\right),
\label{eq:conditonal_pdf}
\end{equation}
where $P(y|x)$ is the intrinsic conditional PDF determined from our Monte Carlo simulations, while $P_{\rm noise}$ describes the measurement uncertainty. 
The observed distribution is thus a convolution between the true conditional PDF, $P\left(y|x\right)$, and $P_{\rm noise}\left(x|\tilde{x}\right)$. 
We assume Gaussian measurement noise in $x$, corresponding to a lognormal uncertainty in $\hat{h}^2_c$.
This conditional distribution can be compared to the marginal distribution for $y$ without incorporating any prior knowledge of $\hat{h}^2_c$.

Fig.~\ref{fig:conditioned_width} shows the resulting widths of the conditional distributions (95\% intervals) at example frequencies of $f = 0.1\, \mathrm{yr}^{-1}$ and $f = 1 \, \mathrm{yr}^{-1}$.
These widths are given as a function of the $\hat{h}^2_c$ measurement uncertainties. The different curves in each panel vary the assumed central values of the noisy $\hat{h}^2_c$ measurements, while the figure legend indicates the percentage of Monte Carlo realizations that yield smaller $\hat{h}^2_c$ values than each case. 

Fig.~\ref{fig:conditioned_width} shows that conditioning reduces the widths of the $\hat{h}^4_c$ and $\hat{C}_{\rm shot}/(4 \pi)$ distributions when the measurement uncertainty is smaller than the intrinsic scatter in $\hat{h}^2_c$. As the uncertainty increases, the widths approach their marginal values. 
Since the $\hat{h}^2_c$ distributions are narrower at low frequencies (e.g., see Fig.~\ref{fig:frequency_dep}), the flattening occurs at lower $\sigma_{\log_{10} \hat{h}^2}$ for $f = 0.1 \, \mathrm{yr}^{-1}$ than for $f = 1 \, \mathrm{yr}^{-1}$. That is, a more precise measurement of $\hat{h}^2_c$ is required at small $f$ to benefit from the conditioning. 
For example, at $f = 0.1 \, \mathrm{yr}^{-1}$ the distributions rise steeply with increasing measurement error and flatten above $\sigma_{\log_{10} \hat{h}^2} \gtrsim 0.1$, whereas the $f = 1 \, \mathrm{yr}^{-1}$ results show a more gradual rise and flatten at $\sigma_{\log_{10} \hat{h}^2} \gtrsim 0.2-0.3$ or so. 
The more stringent measurement requirement at low frequencies may be met in future PTA surveys that determine the characteristic strain with greater precision at the low-frequency end of the observing band.

Fig.~\ref{fig:conditioned_width} shows that precise measurements of $\hat{h}^2_c$ can reduce the 95\% interval widths of the $\hat{h}^4_c$ and $\hat{C}_{\rm shot}/(4 \pi)$ distributions from $\sim 1.5-3$ dex down to $\sim 0.7-0.8$ dex. The exact benefit here depends on frequency, on the central $\hat{h}^2_c$ measurement value, and on whether one considers $\hat{h}^4$ or $\hat{C}_{\rm shot}/(4 \pi)$. Note that in the limit of a perfect $\hat{h}^2_c$ measurement, the widths of the $\hat{h}^4$ and $\hat{C}_{\rm shot}/(4 \pi)$ distributions become identical as these statistics are equivalent in this limit (see, e.g., Eq.~\eqref{eq:cshot_poisson_sampling}). At large $\hat{h}^2_c$ uncertainty, the $\hat{C}_{\rm shot}/(4 \pi)$ distribution is narrower than that of $\hat{h}^4_c$, because the shot-noise in each realization is bounded above by unity. Ultimately, the utility of these two statistics depends not just on the widths of these distributions, but also on their ability to discriminate interesting GWB models. We discuss an example application in \S~\ref{sec:residence_time}. 

This conditioning approach can also be combined with future individual source detections from PTA measurements~\cite{NANOGrav:2023pdq}.
Once bright binaries can be detected individually, they can be excised to measure the strain moments and shot-noise from the remaining unresolved background~\cite{Becsy:2022pnr}. The high-strain tails of the $\hat{h}^2_c$, $\hat{h}^4_c$, and $\hat{C}_{\rm shot}$ distributions will be reduced after this removal, and their breadth will be effectively tightened. This will require further study, however, since the source detection prospects themselves will vary depending on the distribution tails in particular realizations.  

\section{Dependence on SMBHB Merger Models}
\label{sec:model_dependence}

Here, we explore the sensitivity of our results to variations around the assumed
SMBHB merger model. We compare the results in the LM24 and SZQ models and quantify the sensitivity to $z_{\rm min}$ in our fiducial LM24 scenario. We also
investigate alternatives to purely GW-driven inspirals, while illustrating the potential benefit of conditioning the $\hat{h}^4_c$ and $\hat{C}_{\rm shot}/(4 \pi)$ measurements on observational $\hat{h}^2_c$ estimates. 

\subsection{Comparison of Merger Models and $z_{\rm min}$ Sensitivity}
\label{sec:smbh_model}

\begin{figure}
\centering
\includegraphics[width=0.99\linewidth]{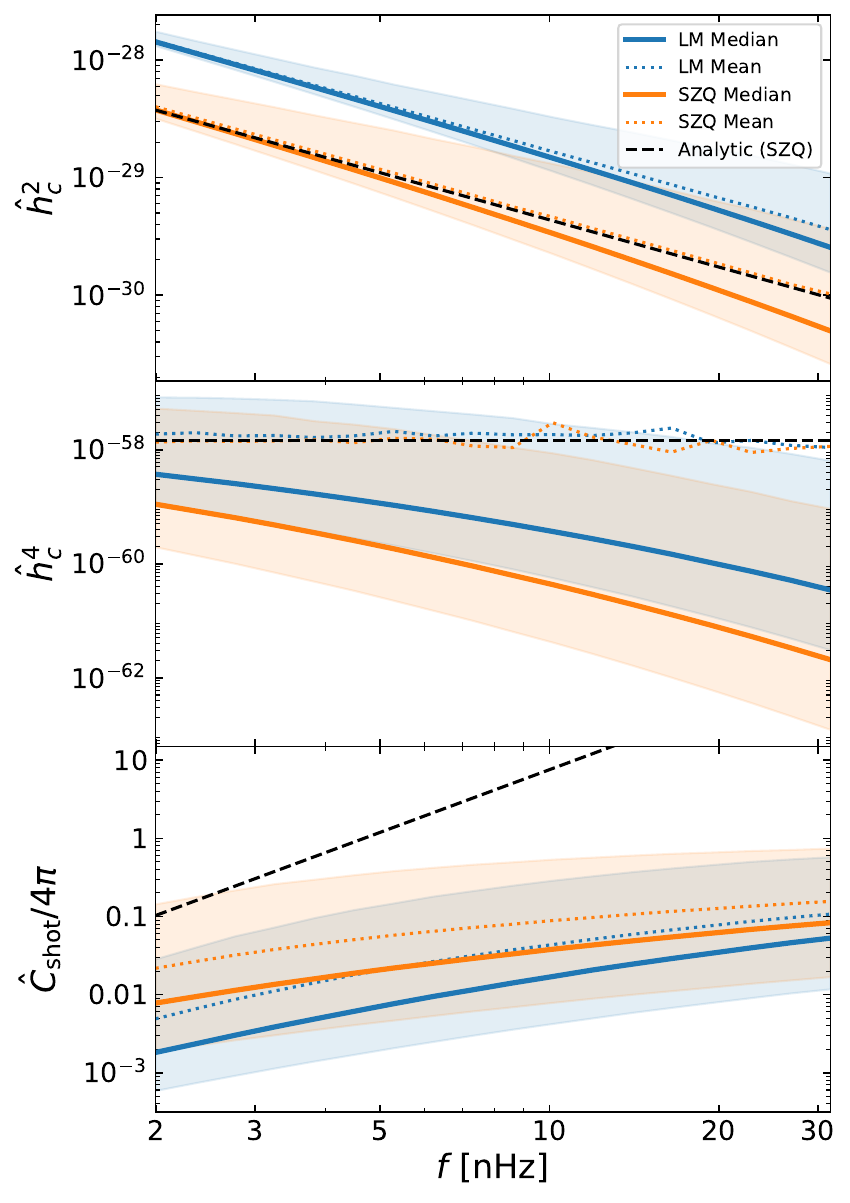}
\caption{Comparison between the strain moments and shot-noise in LM24 (blue) and SZQ (orange). Similar to Fig.~\ref{fig:frequency_dep}, except here we show the corresponding results in SZQ as well. Analytic results are only given in SZQ as the corresponding LM24 predictions are indicated in Fig.~\ref{fig:frequency_dep}.
The $h^2_c(f)$ values in SZQ are smaller than in LM24. At fixed $M_{\rm BH, max} = 10^{10.5} M_\odot$ and $z_{\rm min}=0.05$, the mean $h^4_c(f)$ signal is just slightly higher in LM24, while the median in SZQ falls below that in LM24. The mean and median shot-noise are larger in SZQ, with the median SZQ result being coincidentally similar to the mean in LM24. As in the LM24 model, the moment-based shot-noise prediction in SZQ (black dashed line in the bottom panel) is a significant over-estimate.}
\label{fig:szq_comp}
\end{figure}

Fig.~\ref{fig:szq_comp} compares the LM24 and SZQ results, while fixing $M_{\rm BH, max}=10^{10.5} M_\odot$ and $z_{\rm min} = 0.05$. Although SZQ has a more prominent SMBH mass function tail, exceeding LM24 for $M_{\rm BH,max} \gtrsim 2 \times 10^{10} M_\odot$, it has fewer lower mass black holes (see e.g. Fig. 4 of \cite{Liepold:2024woa}). Consequently, the mean $h^2_c(f)$ values are
$3.6-3.7$ times larger in LM24 than in SZQ at $f \sim 0.1-1 \rm{yr}^{-1}$ ($3-30$ nHz). As discussed in \cite{Liepold:2024woa}, the larger characteristic strain in LM24 is more compatible with current NANOGrav measurements. 
For our fiducial choices of $M_{\rm BH, max}$ and $z_{\rm min}$,
the frequency-independent mean values of $h^4_c(f)$ are just slightly larger, by a factor of $1.3$, in LM24 than in SZQ. This reflects a trade-off between the more prominent SMBH mass function tail in SZQ, and the reduced SZQ mass function at more moderate $M_{\rm BH}$ values. The higher $p_z(z)$ weighting towards low-$z$ also boosts SZQ relative to LM24 \cite{Lin:2026owz}. In sum, for $M_{\rm BH, max} = 10^{10.5} M_\odot$, the shot-noise is larger in SZQ because of the smaller characteristic strain in this model.  
 
Fig.~\ref{fig:szq_comp} reveals that the median $\hat{h}^2_c(f)$ signal in each scenario generally falls beyond the 95\% scatter in the competing model. Therefore, sample variance in $\hat{h}^2_c$ does not seem to be a major obstacle here, and these models should be distinguishable given sufficiently precise $\hat{h}^2_c$ measurements. The middle panel illustrates that although the two models make quite similar predictions for the mean $h^4_c(f)$ signals, the median in SZQ is notably suppressed. This occurs because the median signal reflects typical realizations, while the mean receives important contributions from rare high-strain events in the distribution tail.

Likewise, the median shot-noise signal in the bottom panel of Fig.~\ref{fig:szq_comp} is enhanced in SZQ, exceeding that in LM24 by a factor of $3.5$ at $f \sim 3$ nHz ($0.1 \, \rm{yr}^{-1}$) and $1.6$ at $f \sim$ 30 nHz ($1 \, \rm{yr}^{-1}$). 
As expected from our previous discussion, the mean SZQ shot-noise lies well below the moment-based mean estimate, as indicated by the black-dashed line in the bottom panel of the figure. Nevertheless, the median shot-noise signal in SZQ is still relatively large and encouraging from a detectability point of view, reaching $C_{\rm shot}/(4 \pi) = 0.08$ at $f \sim 1 \, {\rm yr}^{-1}$ (30 nHz).

Note that although the mean $h^4_c$ and $C_{\rm shot}/(4 \pi)$ signals are sensitive to $M_{\rm BH, max}$ in SZQ \cite{Lin:2026owz}, the median should be relatively insensitive to this cutoff. The mean signals receive large weights from the extreme high-mass tail of the SMBH mass function, where the abundance is small enough for only rare realizations to contain such sources. Thus, the median $h^4_c$ and $C_{\rm shot}/(4 \pi)$ signals, shaped by typical realizations, are fairly insensitive to $M_{\rm BH, max}$.

\begin{figure}
\centering
\includegraphics[width=0.99\linewidth]{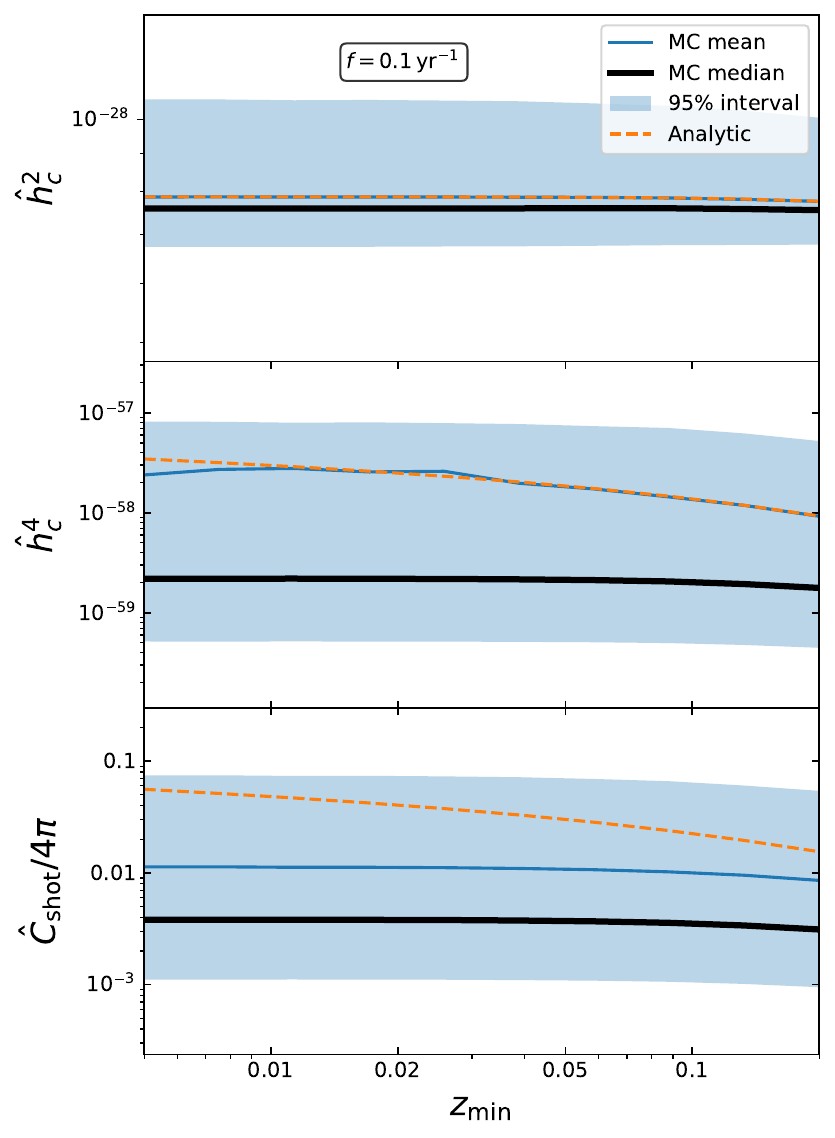}
\caption{Dependence of the strain-moments and shot-noise on $z_{\rm min}$ in the LM24 model at an example frequency of
$f = 0.1 \, \rm{yr}^{-1}$. Although the mean $\hat{h}^4_c$ signal grows with decreasing $z_{\rm min}$, the median signal is insensitive to it. Likewise, the moment-based estimate for the mean $\hat{C}_{\rm shot}/(4 \pi)$ is enhanced at low $z_{\rm min}$, yet neither the mean nor the median shot-noise signals show significant $z_{\rm min}$ dependence across the range shown. 
These trends reflect the importance of the tail of the strain distributions, here from nearby sources, in setting the ensemble-averaged $h^4_c(f)$ values. Typical realizations, on the other hand, are insensitive to rare low $z$ sources. 
}
\label{fig:zmin_dep}
\end{figure}

Fig.~\ref{fig:zmin_dep} further illustrates the dependence on $z_{\rm min}$ in LM24. Our fiducial model adopts $z_{\rm min} = 0.05$, yet this choice is rather arbitrary. Here, we generalize to consider other values of $z_{\rm min}$, in each case adjusting the normalization of $p_z(z)$ so that it integrates to unity. 
Although the mean $h^4_c(f)$ signal is sensitive to nearby sources, as discussed in \cite{Lin:2026owz}, the median signal is insensitive to $z_{\rm min}$ across the range shown, $z_{\rm min} = 0.005 - 0.2$. The shot-noise values are likewise insensitive to $z_{\rm min}$. Future targeted searches for individual sources \citep{NANOGrav:2025gqp} can inform our understanding of the redshift distribution of the nearest sources, while local galaxy catalogs also provide useful guidance \citep{Schutz:2015pza,Mingarelli:2017fbe}.

\subsection{Residence Time Models}
\label{sec:residence_time}

\begin{figure}
\centering
\includegraphics[width=0.99\linewidth]{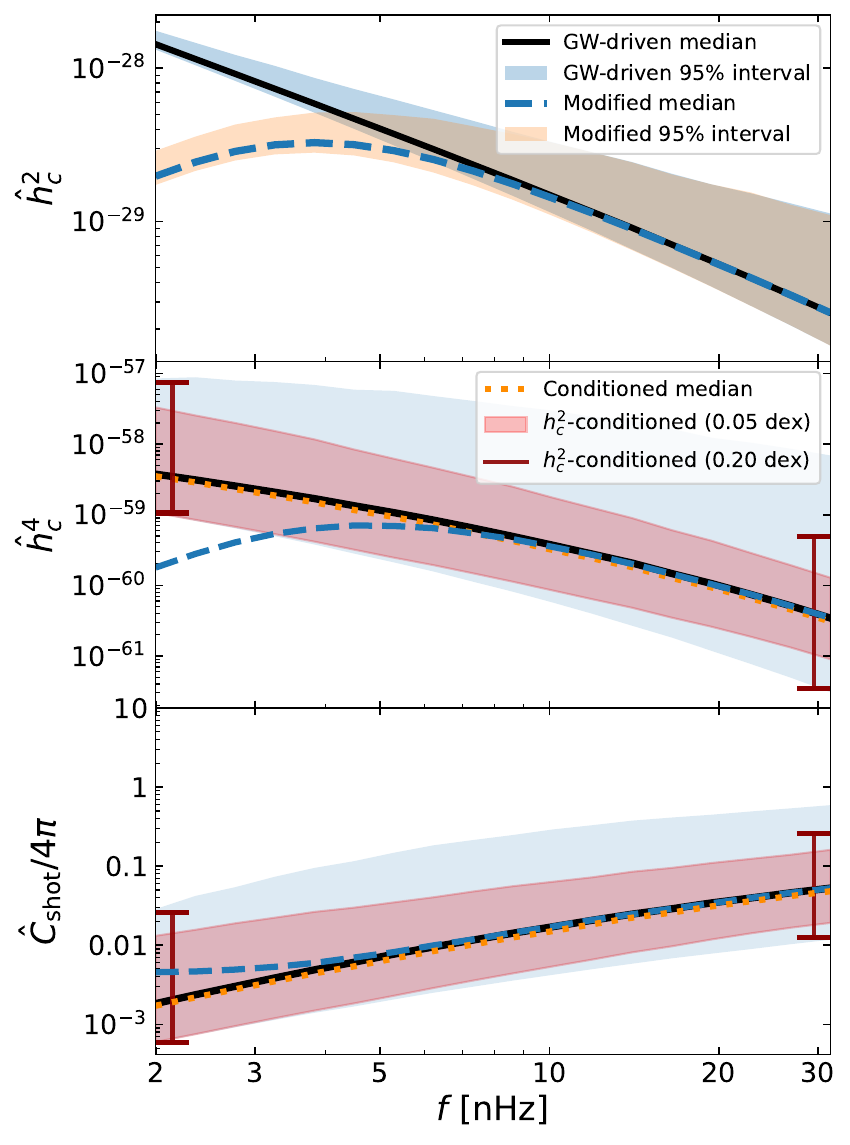}
\caption{The frequency dependence of the strain and shot-noise signals for the fiducial GW-driven model and a stellar-hardening (``Modified'') model.
Stellar hardening reduces the residence times at large separations, suppressing $h^2_c(f)$ (top panel) and $h^4_c(f)$ (middle panel) at low frequencies and increasing shot-noise (bottom panel) due to fewer contributing sources. 
The blue shaded regions show the 95\% realization-to-realization scatter in the GW-driven scenario, while the orange shaded region in the $h^2_c(f)$ panel shows the same for the stellar-hardening case.  The red shaded regions in the middle and bottom panel indicate the 95\% intervals after
conditioning on a 0.05 dex measurement of $h^2_c(f)$. The red error bars, at the frequency edges, show the 95\% intervals for a 0.20 dex measurement.
The dark orange dotted line shows the median after conditioning, which is barely changed. 
}
\label{fig:stellar_hardening}
\end{figure}

The shot-noise and $h^4_c(f)$ signals depend on the SMBHB residence times, and  measurements of these quantities may therefore help constrain residence time models and provide consistency tests beyond inferences from $h^2_c(f)$ alone \cite{Sato-Polito:2023spo,Lamb:2024gbh,Lin:2026owz}. At the low frequency end of the PTA band, the residence times may be set by three-body interactions between the SMBHBs and surrounding stars and by gas dynamics. Hence, PTA measurements may shed light on the environmental interactions between SMBHBs and their host galaxies that drive the black holes to merger, provided  that environmental effects remain important at PTA-relevant orbital separations.
However, the large realization-to-realization scatter in
$\hat{h}^2_c$, $\hat{h}^4_c$, and $\hat{C}_{\rm shot}/(4 \pi)$ may potentially prevent distinguishing interesting model variants from such measurements.

In order to explore this quantitatively, we construct an alternate toy model for SMBHB residence times, intended to mimic the possible effects of stellar-hardening in accelerating the early phases of the inspiral process. Specifically, following references~\cite{Sampson:2015ada,Sato-Polito:2023spo} we suppose that stellar-hardening leads to a faster inspiral evolution with $d\ln f_r/dt_r|_{\rm tot} = d \ln f_r/dt_r|_{\rm GW} \, + d \ln f_r/dt_r|_{\rm SH}$, where $d \ln f_r/dt_r|_{\rm GW}$ denotes the GW-driven inspiral and  $d \ln f_r/dt_r|_{\rm SH}$ is from stellar-hardening. The resulting residence time can be described by \cite{Sampson:2015ada,Sato-Polito:2023spo}:
\begin{equation}
\left.\frac{dt_r}{d \ln f_r}\right|_{\rm tot} = \left.\frac{dt_r}{d \ln f_r}\right|_{\rm GW} \times \, \left[\frac{1}{1 + \left(f_b/f_r \right)^\kappa} \right],
\label{eq:dtr_mod}
\end{equation}
where $f_r$ is the source-frame frequency, $\kappa=10/3$ for stellar-hardening, and $f_b$ is a bend frequency \cite{Sampson:2015ada}. Hence, at $f_r \gg f_b$, one recovers the GW-driven case with $dt_r/d \ln f_r \propto f^{-8/3}_r$ while at $f_r \ll f_b$, the evolution is stellar-hardening-dominated with $dt_r/ d \ln f_r \propto f^{\kappa - (8/3)}_r \propto f^{2/3}_r$.

A simplifying feature of this model is the assumption of a SMBHB mass-independent bend frequency, $f_b$. A more detailed calculation would account for mass and mass-ratio dependence in $f_b$, since these and environmental parameters shape the competition between GW emission and stellar-hardening which set $f_b$.  
As $h^4_c(f)$ receives most of its contributions from higher mass SMBHBs than $h^2_c(f)$ \cite{Lin:2026owz}, the mass dependence of $f_b$ might impact the relative trends of $\hat{h}^2_c$, $\hat{h}^4_c$, and $\hat{C}_{\rm shot}/(4 \pi)$ with frequency. For simplicity, here we nevertheless adopt the mass-independent $f_b$ approximation, taking
$f_b = 5$ nHz and $\kappa = 10/3$.

We construct Monte Carlo simulations in this stellar-hardening model (denoted ``Modified'') and contrast the results with our fiducial GW-driven scenario in Fig.~\ref{fig:stellar_hardening}. As expected, $h^2_c(f)$ declines below the bend frequency, since the SMBHBs are evolving more quickly through the early stages of inspiral and fewer sources contribute to the strain signal. At $f \ll f_b$, the scaling becomes $h^2_c(f) \propto (dt_r/d \ln f) \times \, h^2_s(f) \propto f^{2}$, reflecting the shortened residence time in the stellar-hardening regime.

The suppression in $h^2_c(f)$ is large compared to the realization-to-realization scatter, making it a powerful discriminator between the models. The behavior of $h^4_c(f)$ is qualitatively similar, and despite its broader intrinsic scatter, the modified median falls below the fiducial 95\% interval. 
For comparison, we also calculate the conditional $\hat{h}^4_c$ and $\hat{C}_{\rm shot}/(4 \pi)$ distributions discussed in \S~\ref{sec:conditioned}. The conditional distribution calculations assume that the central value of $\hat{h}^2_c$ matches
the fiducial model median at each frequency, with a frequency-independent measurement uncertainty of $0.05$ dex. 
As expected from Fig.~\ref{fig:conditioned_width}, this significantly reduces the effective scatter. For comparison, the red error bars in the figure also illustrate
the results after conditioning with a larger $0.20$ dex measurement uncertainty. For visual clarity, we show these results only at the edges of the PTA frequency band. In this case, the
$\hat{h}^4_c$ and $\hat{C}_{\rm shot}/(4 \pi)$ bands tighten near the high frequency end where the intrinsic spread in $\hat{h}^2_c$ is greater, while there is little benefit at low frequencies where the intrinsic $\hat{h}^2_c$ distribution is narrower. This follows our expectations from Fig.~\ref{fig:conditioned_width}.
Note that the median $\hat{h}^4_c$ and $\hat{C}_{\rm shot}/(4 \pi)$ values are barely changed ($\lesssim 10\%$) after conditioning (see the dark orange dotted curves in the figure). 

The reduced scatter from  conditioning is not crucial here, however, as the stellar-hardening case falls below even the marginal $\hat{h}^4_c(f)$ scatter. Put differently, if the true model is the stellar-hardening scenario, one should be able to rule out the GW-driven model from $\hat{h}^2_c(f)$ observations with sufficiently small measurement errors, while $\hat{h}^4_c(f)$ may provide a consistency test.

In contrast, the median shot-noise signal shows only a modest increase due to the reduced source abundance in the modified model and remains within the conditional scatter. Since the shot-noise involves the ratio of $\hat{h}^4_c$
and $(\hat{h}^2_c)^2$ (Eq.~\eqref{eq:cshot_poisson_sampling}), the suppression in numerator and denominator partially cancels. As a result, shot-noise alone is a less effective discriminator between these models. 
This highlights the potential benefits of combining measurements of $\hat{h}^2_c$, $\hat{h}^4_c$, and $\hat{C}_{\rm shot}/(4 \pi)$.

\section{Conclusions}
\label{sec:conclusions}
This work has characterized the probability distribution of shot-noise amplitudes across sky realizations as a function of observing frequency, using Monte Carlo simulations drawn primarily from the LM24 SMBHB merger model.  
Our main conclusions are as follows. First, analytic estimates of the mean shot-noise based on the ratio of moments, $\propto \langle h^4 \rangle_\Omega/\langle h^2 \rangle_\Omega^2$, can differ dramatically from the ensemble-averaged shot-noise.  In our fiducial model, this discrepancy exceeds two orders of magnitude by $f = 1 \, \mathrm{yr}^{-1}$. Second, the shot-noise distribution is highly skewed: the most probable and median shot-noise lie below the mean values by factors of $\sim 2-3$ across the PTA band. The distribution is also broad, spanning a 95\% interval of 1.7 dex, reflecting the strong influence of rare bright sources. Third, the shot-noise amplitudes remain substantial in typical realizations. In our fiducial model, the median shot-noise grows from approximately $\hat{C}_{\rm shot}/(4 \pi) = 3 \times 10^{-3}$ at $f \sim 3$ nHz to about $\hat{C}_{\rm shot}/(4 \pi) = 0.05$ at $f \sim 30$ nHz,  scaling as 
$\hat{C}_{\rm shot} \propto f^{1.1-1.2}$ in a GW-driven scenario. Although below 
both moment-based estimates and the ensemble means, these median amplitudes remain within the prospective reach of future PTA anisotropy measurements, with the 
current 95\%
upper limits already at the $\hat{C}_{\rm shot}/(4 \pi) \leq 0.20$ level \cite{NANOGrav:2023tcn}.

Fourth, we characterize the related distributions of $h^2_c(f)$ and $h^4_c(f)$. 
The $h^4_c(f)$ statistic, which does not involve normalization by $1/\left(\hat{h}^2_c\right)^{2}$, provides a complementary diagnostic of the source populations. 
Its distribution is even broader than that of the shot-noise, spanning  a 95\% interval of $\sim$ 3 dex. While the ensemble average of $h^4_c(f)$ is independent of frequency in a GW-driven scenario, the median declines with frequency, 
reflecting the decreasing source abundance and the growing importance of rare high-strain sources. We find that conditioning on measurements of $\hat{h}^2_c(f)$ can reduce the effective width of the shot-noise and $\hat{h}^4_c(f)$ distributions. For sufficiently precise $\hat{h}^2_c(f)$ measurements, the shot-noise distribution can narrow by up to 1 dex, while the $\hat{h}^4_c(f)$ scatter can tighten by more than 2 dex. 

Finally, we explore the sensitivity of these results to aspects of our SMBHB population models. We begin by considering variations around our fiducial LM24 SMBH mass function and the sensitivity to the low-redshift end of the merger rate distribution. We find that the shot-noise is somewhat larger in the SZQ SMBHB merger model, and that the median signals are insensitive to $z_{\rm min}$.
We then investigate an alternative residence time model intended to mimic the role of stellar-hardening in accelerating early phases of the inspiral process. We find that this leads to distinctive declines in $\hat{h}^2_c(f)$ and $\hat{h}^4_c(f)$ at low frequencies. The shot-noise, however, is less impacted by these effects as it depends on a ratio of strain moments and partial cancellations occur.  

Our results demonstrate that shot-noise amplitudes follow broad distributions and cannot be characterized by ensemble-averaged quantities alone. Interpreting PTA anisotropy measurements therefore requires modeling the full PDFs of the relevant observables. In future work, it will be valuable to extend this treatment to account for PTA frequency-weighting schemes. That is, one should average the differential quantities considered here, defined per $d \ln f$, over the finite frequency bins appropriate for the PTA data set. It will also be important
to explore the covariance of shot-noise estimates across different multipoles, and to construct mock sky maps that capture the non-Gaussian angular structure of the signal. Ultimately, properly characterized shot-noise PDFs will be an important ingredient for future PTA parameter inference pipelines.

\acknowledgments

We thank Stephen Taylor and Chiara Mingarelli for helpful discussions. 
M-X.L. is supported in part by the Canadian Institute for Theoretical Astrophysics (CITA) National Fellowship. C.-P. M. acknowledges the support of NSF AST-2307719.

\bibliographystyle{apsrev4-1}
\bibliography{ref}

\end{document}